\documentclass[12pt]{article}
\usepackage{hyperref}
\usepackage{fancyvrb}

\begin{document}

\title{What can we learn from NJL-type models about dense matter?}

\author{Michael Buballa\\
Theoriezentrum, Institut f\"ur Kernphysik, TU Darmstadt\\
D-64289 Darmstadt, Germany}

\date{}

\maketitle

\begin{abstract}
The merits and limitations of the Nambu--Jona-Lasinio model as a model for strong interactions at nonzero density
are critically discussed. 
We present several examples, 
demonstrating that, while in general the results should not be trusted quantitatively,
the NJL model is a powerful theoretical tool for getting new insights 
and ideas about the QCD phase diagram and the dense-matter equation of state. 
\end{abstract}

\section{Introduction}

In 1961, Nambu and Jona-Lasinio published two seminal papers on a ``Dynamical Model of Elementary Particles Based on an 
Analogy with Superconductivity,'' \cite{Nambu:1961tp,Nambu:1961fr},
now known as the Nambu--Jona-Lasinio (NJL) model.
Originally invented as a model for point-like nucleons, 
after the advent of QCD the NJL model was reinterpreted as a schematic model
for quarks, interacting by four-point vertices rather than by the exchange of gluons. 
Subsequently the model was extended from two to three quark flavors
and applied to study effects of nonzero temperature and chemical potential
as well as color superconductivity (for reviews, see \cite{Vogl:1991qt,Klevansky:1992qe,Hatsuda:1994pi,Buballa:2003qv}).
More recently features of Polyakov-loop dynamics have been added to the model by coupling the quarks to 
a background temporal gauge field with a phenomenological potential (PNJL model)
\cite{Meisinger:1995ih,Fukushima:2003fw,Megias:2004hj,Ratti:2005jh}.

The ground-breaking achievements of the original NJL papers were to explicitly demonstrate 
how the spontaneous breaking of chiral symmetry in a quantum-field theoretical context leads to the dynamical generation 
of fermion masses, while at the same time there appears a massless mode (``Nambu-Goldstone boson'') in the 
quark-antiquark scattering matrix, which can be identified with the pion.  
However, despite this indisputable success (culminating in the 2008 Nobel prize awarded to Nambu  ``for the discovery 
of the mechanism of spontaneous broken symmetry in subatomic physics'')  one may ask why
we should still use a model after QCD was established as the theory of the strong interaction.
Of course, model calculations are in general much simpler than QCD calculations. 
But to what extent can we trust the results? In particular we have to face the following problems:
\begin{itemize}
\item
The essential feature of the NJL model which motivates its use 
as an effective model of QCD is the fact that the two share the same global
symmetries. However, the symmetries alone do not uniquely fix the interaction. 
There could be (infinitely) many possible interaction terms, not only containing four-point but also higher
$2n$-point vertices (see, e.g., Ref.~\cite{Benic:2014iaa} and references therein), and thus many parameters.

\item
In principle, these vertices should be derivable from QCD by integrating out the gluonic degrees of freedom.
However, this procedure would lead to density dependent coupling constants,
while in actual NJL-model calculations the model parameters are typically fitted to vacuum observables and 
then kept unchanged in the medium.

\item
As far as symmetries are concerned, there are model independent theorems, which, 
if not spoiled by an improper approximation, are reproduced by the model. 
But those we know anyway while for non-universal properties it is not clear whether the results obtained in the model 
agree with those in QCD.
\end{itemize}
In addition, the NJL model has the well-known shortcomings that it is non-renormalizable and has no 
confinement,\footnote{In the PNJL model confinement is  {\it statistically} realized, meaning that the quark contribution to 
the pressure is suppressed at low temperatures. However, this does not prevent the unphysical decay of mesons into quarks 
and antiquarks in the model~\cite{Hansen:2006ee}.}
which could both lead to artifacts.

There are nevertheless situations where QCD-inspired models in general and specifically the NJL model
can be very useful. 
``Model independent'' predictions are sometimes based on unrealistic assumptions, e.g., Taylor expansions in parameters
which are not really small. Such cases can be uncovered by specific model calculations. 
Sometimes models can also be used to test ideas and techniques used in other frameworks.
Most importantly, however, models can be employed for exploratory studies in order to identify interesting problems, 
worthwhile to be studied more seriously.  

In the following these statements will be illustrated by specific examples
related to the QCD phase diagram and the dense-matter equation of state.

\section{Phase diagram at nonzero temperature and density}

Despite tremendous theoretical and experimental efforts,
the exact phase structure of QCD as a function of temperature and baryon chemical potential $\mu_B \equiv 3\mu$ is still 
unresolved to a large extent~\cite{BraunMunzinger:2008tz,Fukushima:2010bq}.  
While at $\mu = 0$ QCD can be solved on the lattice by standard Monte-Carlo methods, 
this is prevented at $\mu \neq 0$ by the so-called sign problem. 
Our current picture in this regime is therefore mainly based on model calculations, with the NJL model playing a pioneering 
role: 

In 1989 Asakawa and Yazaki presented an NJL-model calculation of the $T$-$\mu$ phase diagram~\cite{Asakawa:1989bq}. 
At low temperatures but high chemical potential they found a first-order chiral phase transition, 
while at low $\mu$ the transition is a crossover, in agreement with today's lattice QCD results.
As a consequence there is a critical point where the first-order phase boundary ends.
Although not much attention was paid to this fact at the beginning, this changed dramatically after it was argued that 
the critical endpoint (CEP) could have observable consequences~\cite{Stephanov:1998dy}. 
Today the search for the CEP is the main goal of the beam-energy scan at RHIC~\cite{Odyniec:2013aaa}
and of the future projects at NICA~\cite{NICA} and CBM at FAIR~\cite{Friman:2011zz}.

To my knowledge the NJL-model calculation of Ref.~\cite{Asakawa:1989bq} was the first prediction of the CEP.  
On the other hand, it was already demonstrated in that reference that its exact position depends on the choice of the
model parameters. 
For instance the CEP can be moved around considerably by varying the strength in the vector channel
or of the chiral anomaly (parametrized by a six-point interaction in the three-flavor model)~\cite{Fukushima:2008wg}. 
Indeed, already for rather moderate values of the vector coupling, the first-order phase boundary (and hence the CEP) 
disappears completely. 

We can thus conclude that the NJL model (like other models) cannot predict the position of the CEP and not even tell
whether it exists. However, it gave the first hint for its possible existence and in this way inspired experimental searches
and more serious theoretical investigations.
In particular there are now various  works which try to identify the CEP directly starting from QCD,
both, within functional methods which do not have a sign problem (like truncated Dyson-Schwinger 
equations~\cite{Fischer:2014mda}) 
and on the lattice, trying to circumvent the sign problem in some way~\cite{deForcrand:2010ys}.
For example, one method to get information about the $\mu \neq 0$ regime by lattice calculations 
is to perform a Taylor expansion of the pressure in powers of $\mu$, evaluating the coefficients at $\mu = 0$. 
The power of this method can in turn be tested within models which do not have a sign problem and thus allow for a direct 
comparison of the Taylor-expanded pressure with the exact model results at $\mu \neq 0$.
Such test have been performed in the NJL model~\cite{Buballa:2008ru}
as well as in the Polyakov-loop extended quark-meson model~\cite{Karsch:2010hm},
revealing that the number of expansion coefficients required for the detection of a CEP located at $\mu/T >1$
would be far beyond the present state of the art.

In most studies of the QCD phase diagram it is tacitly assumed that the chiral condensate, i.e., 
the order parameter for chiral-symmetry breaking is  spatially homogeneous. 
Allowing for spatially varying condensates, however, it turns out that in the NJL model there is a region where such
an inhomogeneous condensate is energetically favored over homogeneous or vanishing 
condensates~\cite{Nakano:2004cd,Nickel:2009ke}.
In particular it was found that for the standard NJL Lagrangian the inhomogeneous phase covers the entire first-order
phase boundary which is obtained in the case when only homogeneous phases are considered~\cite{Nickel:2009wj}.
Moreover, the inhomogeneous phase turned out to be very robust against various model extensions, like
including strange quarks~\cite{Moreira:2013ura}, isospin asymmetries~\cite{Nowakowski:2015ksa}, 
magnetic fields~\cite{Frolov:2010wn} or, most notably, vector interactions~\cite{Carignano:2010ac}
(for a review, see Ref.~\cite{Buballa:2014tba}).
Again, the model cannot be used to prove the existence of such phases in QCD, 
but in the same way as the model prediction of a CEP,  the possibility of an inhomogeneous phase
should seriously be considered and deserves more thorough investigations.
Indeed, inspired by the NJL model results, inhomogeneous phases have also been studied within Dyson-Schwinger QCD,
where qualitatively similar results have been found~\cite{Muller:2013tya}.

\section{Equation of state for compact stars}

Neutron stars can reach densities of several times nuclear-matter densities in their centers. 
Under these conditions it has been argued long time ago that matter could be 
deconfined~\cite{Ivanenko:1965dg,Itoh:1970uw}, so that ``neutron stars'' would 
in fact be hybrid stars with an outer hadronic part and a quark matter core. 
Although this idea has been challenged by the recent discovery of two compact stars with masses of about 
$2 M_\odot$ (where $M_\odot$ is the mass of the sun)~\cite{Demorest:2010bx,Antoniadis:2013pzd},
the question whether or not there are deconfined quarks at the centers of compact stars is still 
open~\cite{Alford:2006vz,Buballa:2014jta}.
Here the problem is again that QCD at $\mu \neq 0$ cannot be studied on the lattice, and that therefore the QCD 
equation of state (EoS) at nonzero density is largely unknown. In this situation one often 
starts from two independent EoSs, a phenomenological hadronic one and a quark-matter one and constructs a phase transition
between them by comparing their pressure at given chemical potential. 

For the quark-matter part the most common choice are MIT bag-model EoSs, but more recently 
NJL-model EoSs have also gained popularity.
One reason is that the critical chemical potential of the phase transition depends sensitively on the bag constant, 
which is a largely unconstrained parameter in the bag model, while in the NJL model it is dynamically generated 
as the pressure difference between the vacuum states with spontaneously broken and unbroken chiral symmetry. 
Moreover, the NJL model allows for a straightforward incorporation of color superconductivity~\cite{Buballa:2003qv}.
Yet, as pointed out earlier, the NJL model has many parameters as well. 
Indeed, while early studies mostly disfavored the presence of quark matter in neutron 
stars~\cite{Schertler:1999xn,Baldo:2002ju}, 
later analyses succeeded in getting solutions with a quark-matter core, 
simultaneously reaching maximum masses above $2 M_\odot$,
by choosing relative large couplings in the vector and diquark channels~\cite{Klahn:2006iw}.
Hence, the NJL model (in combination with a hadronic model) can serve as a counter-example against the claim 
that the detection of compact stars with $2M_\odot$ already rules out the presence of a quark-matter 
core~\cite{Ozel:2006km}.
On the other hand, we conclude again that it cannot make qualitative or even quantitative predictions about its existence.

In fact, it is not even clear, whether resorting to the dynamically generated bag pressure of the NJL model really makes sense
when combining it with a hadronic model.
Basically it means that the NJL model is taken seriously in vacuum and at high densities, but not in the hadronic phase in between.
Some authors therefore introduced an additional bag constant by hand, which is supposed to 
account for confinement effects and other uncertainties~\cite{Pagliara:2007ph,Lenzi:2012xz}.
However, it is then even more questionable to fix the NJL-model parameters in vacuum, and one may ask why the NJL model
should be used at all. (After all, the most important feature of the NJL model is its nontrivial vacuum.)
To my opinion, the only convincing way to ultimately avoid these problems is to describe quark and hadronic phase
 in a single framework, e.g.,  construct nucleons and nuclear matter within the NJL model as well.
Some steps in this direction have been made in Refs.~\cite{Rezaeian:2005nm,Lawley:2006ps,Wang:2010iu}.
As an alternative approach it might also be worthwhile to revisit the description of baryons as chiral 
solitons~\cite{Alkofer:1994ph,Christov:1995vm}
 and investigate their relation to the inhomogeneous phases discussed above.

%



\end{document}